\documentclass[a4paper,fleqn]{cas-sc}

\usepackage{float}
\usepackage{subcaption}
\usepackage{placeins}
\usepackage{pifont}
\usepackage{longtable}
\usepackage{enumitem}
\usepackage[authoryear]{natbib}

\def\tsc#1{\csdef{#1}{\textsc{\lowercase{#1}}\xspace}}
\tsc{WGM}
\tsc{QE}

\begin{document}
\let\WriteBookmarks\relax
\def\floatpagepagefraction{1}
\def\textpagefraction{.001}
\let\printorcid\relax 


\shortauthors{Ding et al.}

\title[mode = title]{How built environment shapes cycling experience: A multi-scale review in historical urban contexts}

\author[1,2]{Haining Ding}[style=chinese]

\cormark[1] 
\credit{}

\author[1,2]{Chenxi Wang}

\credit{}

\author[1,2]{Michal Gath-Morad}
\cormark[1] 
\credit{}

\address[1]{Cambridge Cognitive Architecture, Department of Architecture, University of Cambridge, UK}
\address[2]{NeuroCivitas Lab for NeuroArchitecture, Centre for Research in the Arts, Social Sciences and Humanities (CRASSH), University of Cambridge, UK}

\cortext[1]{Corresponding author. Cambridge Cognitive Architecture, Department of Architecture, University of Cambridge, UK. \\
\hspace*{2em} \textit{Email Address}: hd518@cam.ac.uk (H. Ding), mg2068@cam.ac.uk (M. Gath-Morad).}

\begin{abstract}
Understanding how built environments shape human experience is central to designing sustainable cities. Cycling provides a critical case: it delivers health and environmental benefits, yet its uptake depends strongly on the experience of cycling rather than infrastructure alone. Research on this relationship has grown rapidly but remains fragmented across disciplines and scales, and has concentrated on network-level analyses of routes and connectivity. This bias is especially problematic in historical cities, where embedding new infrastructure is difficult, and where cycling experience is shaped not only by spatial form but also by how cyclists perceive, interpret, and physically respond to their environment --- through psychological factors such as safety and comfort, physiological demands such as stress and fatigue, and perceptual cues in the streetscape. We systematically reviewed 68 studies across urban planning, transportation, behavioural science, neuroscience, and public health. Two scales of analysis were identified: a macro scale addressing the ability to cycle and a micro scale addressing the propensity to cycle. Methods were classified into objective and subjective approaches, with hybrid approaches beginning to emerge. We find a persistent reliance on objective proxies, limited integration of subjective accounts, and insufficient attention to the streetscape as a lived environment. Addressing these gaps is essential to explain why environments enable or deter cycling, and to inform the design of cities that support cycling as both mobility and lived experience.
\end{abstract}



\begin{keywords}
Cycling experience \sep 
Built environment \sep 
Historical urban contexts \sep
Systematic review
\end{keywords}

\maketitle


\section{Introduction}

Research has consistently confirmed that the built environment shapes cycling behaviour through spatial configurations such as network connectivity, infrastructural coherence, and route directness \citep{chen_quantify_2025, brauer_characterizing_2021, ding_towards_2021, wang_relationship_2020, weikl_data-driven_2023}. Yet cycling is not merely a matter of provision. It is a situated and embodied practice, where cyclists engage with the environment through perceptual, psychological, and physiological processes. Unlike enclosed modes of transport, cyclists are fully exposed to the street. They navigate its textures, interpret its cues, and absorb its frictions. This exposure renders cycling highly sensitive to the spatial and affective qualities of urban form, with direct implications for the design of sustainable mobility systems \citep{basil_exploring_2023, aldred_cycling_2017, battiston_revealing_2023}.

Despite growing interest in this dimension, existing research remains fragmented across disciplines and spatial scales. At the macro scale, cycling tends to be modelled through objective attributes, such as accessibility and directness, often using data-driven indicators at regional or neighbourhood levels \citep{ciris_investigating_2024, rerat_build_2024}. These studies emphasise the ‘ability to cycle’, yet offer limited insight into how that space is actually perceived. At the micro scale, attention shifts to the ‘propensity to cycle’, explored through subjective dimensions such as safety, comfort, and attractiveness \citep{ahmed_bicycle_2024, biassoni_choosing_2023, bialkova_how_2022}. While these perspectives acknowledge lived experience, they are often disconnected from spatial modelling. Recent hybrid approaches have begun to bridge this divide, incorporating tools such as computer vision, street-level imagery, and shared mobility data \citep{felix_reproducible_2025, zare_simple_2024, garrido-valenzuela_where_2023, ito_translating_2024, wang_planning_2023, gong_deciphering_2024}. Still, few reviews have systematically examined how cycling experience is conceptualised and measured across these diverse approaches \citep{schon_scoping_2024, yang_key_2023}.

This limitation is particularly pronounced in historical urban contexts, where formal constraints restrict infrastructural transformation. In such cities such as Cambridge, cycling experience is not determined solely by infrastructure, but by how space is seen, interpreted, and inhabited  \citep{nazemi_studying_2021, ghekiere_assessing_2015}. Visual cues such as openness, greenness, and spatial continuity shape how streets are perceived and navigated, influencing both psychological ease and physical effort \citep{gath-morad_beyond_2022}. These elements are rarely foregrounded in large-scale models, yet they play a crucial role in environments where conventional interventions are infeasible.

In this paper, we provide a structured review of how cycling experience has been conceptualised and studied in historical urban contexts, with attention to spatial scales, methodological approaches, and disciplinary orientations. We synthesise recent advances in evaluating the built environment as a lived context for cycling, and identify key limitations, research gaps, and emerging directions. To the extent of our knowledge, this is the most comprehensive and systematically organised review on this topic. The review is guided by two questions:

\begin{enumerate}[label=\roman*.]
    \item What factors in historical urban contexts determine cycling experience across urban scales?
    \item Which methods are applied to evaluate cycling experience in various urban scales in historical urban contexts?
\end{enumerate}

\section{Related studies}

To the best of our knowledge, four review papers published in international scientific outlets are most closely related to our study.

\citet{kellstedt_scoping_2021} and \citet{ahmed_bicycle_2024} provide structured overviews of bikeability indices, focusing on how built environment features are operationalised. Both reviews foreground macro-scale variables such as connectivity, route directness, and land-use diversity, drawing on planning and transportation literature. However, the role of perception is largely treated as a derived outcome, with limited attention to the experiential dynamics of cycling or their spatial triggers.

\citet{kalra_methods_2023} concentrate on subjective experience, compiling methods used to capture cyclists’ psychological and emotional states. Their scope spans self-reports, biosensors, and mobile sensing. While valuable in identifying affective dimensions, the review remains methodologically isolated from spatial frameworks. It does not explore how built environments contribute to or shape these experiences across different urban settings.

\citet{ito_understanding_2024} examine how visual cues in the built environment are quantified using computational methods. Although the review introduces relevant perceptual indicators such as greenness and enclosure, its focus lies in method development rather than cycling. The absence of a mobility lens, particularly for exposed modes like cycling, limits its applicability to transport-oriented design questions.

In our review, we build on these insights by tracing how cycling experience has been conceptualised and assessed across spatial scales, from infrastructural provision to streetscape perception. We focus specifically on historical urban contexts, where the constraints of form render perception not a residual layer but a central mechanism through which space is interpreted and navigated.

\section{Methods}

This systematic review investigates how built environment characteristics influence cycling experience across multiple urban scales. The review process follows the PRISMA guidelines for systematic reviews \citep{page_prisma_2021}, including structured stages of literature search, screening, and data extraction. To improve the efficiency and consistency of relevance screening, we employed ASReview, a machine learning-based tool designed to prioritise studies using active learning \citep{van_de_schoot_open_2021}.

\subsection{Search strategy}

A structured literature search was conducted on 9 November 2024 across four academic databases: Web of Science (WOS), Scopus, PubMed, and Open Wiley Library. The search aimed to identify peer-reviewed studies published since 2014 that examined built environment determinants of cycling experience using valid methods. The selection of search terms was designed to capture both physical and perceptual aspects of cycling across urban scales, while excluding irrelevant uses of the word ‘cycling’ in other disciplines such as biology or chemistry.

\begin{figure}[ht]
\centering
\includegraphics[width=1\linewidth]{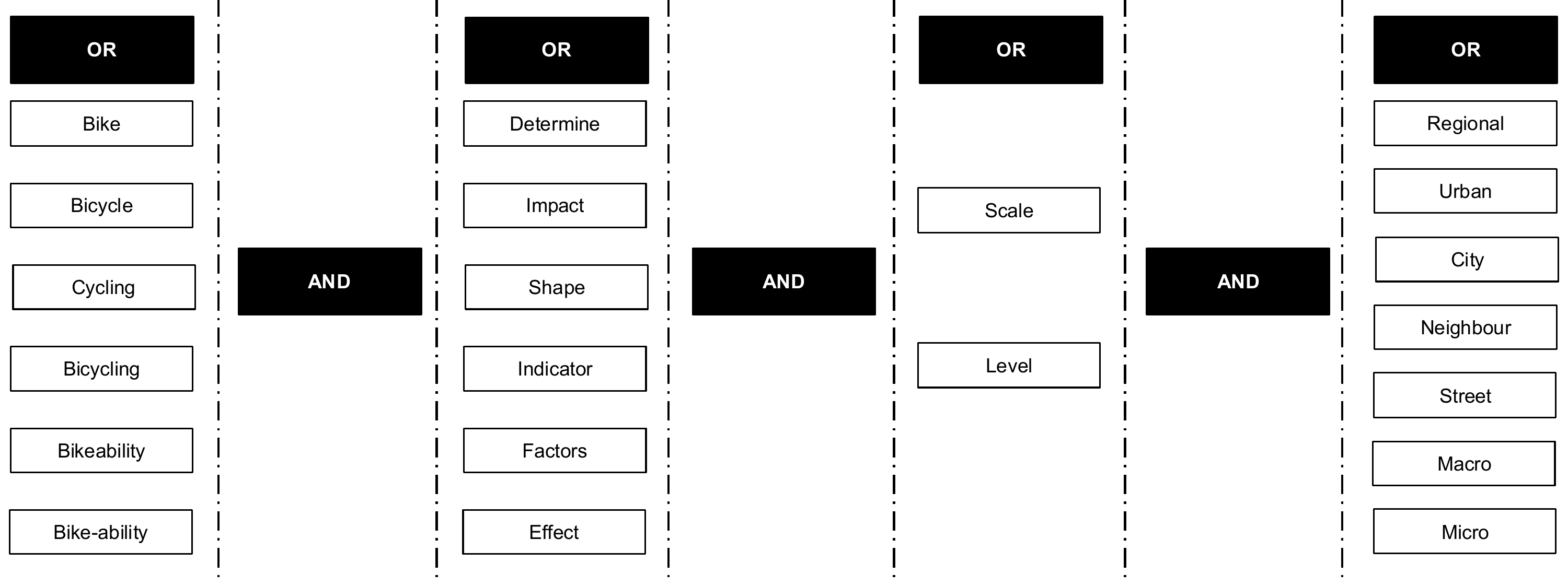}
\caption{Search Terms Applied}
\label{fig 1:Search Terms Applied}
\end{figure}

Search strings applied three layers of conditions (Figure \ref{fig 1:Search Terms Applied}). First, publication titles were required to contain cycling-related terms using the Boolean logic: \texttt{(bike OR bicycle OR cycling OR bicycling OR bikeability OR "bike-ability")}. Second, abstracts needed to include at least one term relating to explanatory mechanisms or influences, such as \texttt{(determinant OR determine OR impact OR shape OR indicator OR factor OR effect)}. Third, the abstracts also had to reference spatial scale explicitly or implicitly, using terms like \texttt{(scale OR level)} in combination with spatial contexts, including \texttt{(urban OR regional OR city OR street OR neighbourhood OR "large scale" OR "micro scale" OR "macro scale")}. Boolean operators (AND/OR) were used to combine these conditions across fields, with adjustments made to match the syntax of each database.

\subsection{Literature selection}

\begin{figure}[ht]
\centering
\includegraphics[width=1\linewidth]{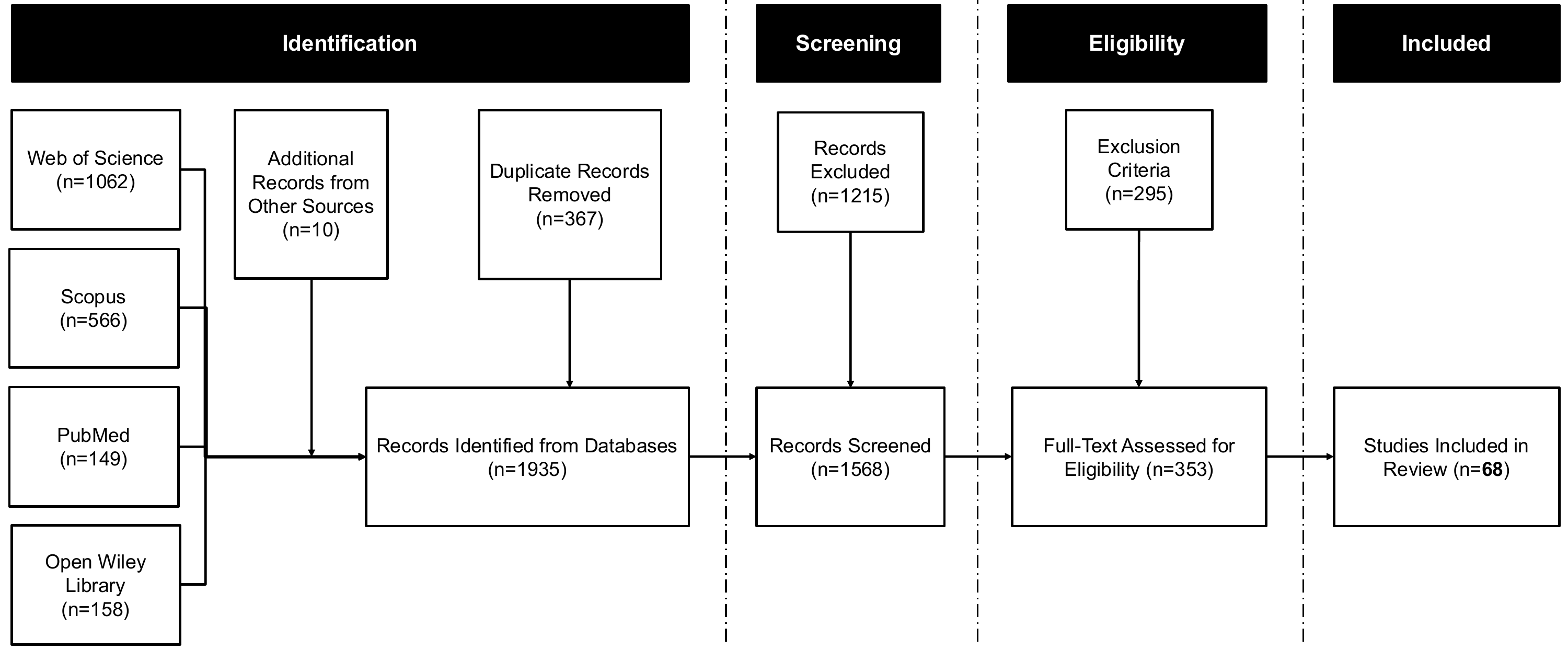}
\caption{PRISMA Flowchart for Literature Review}
\label{fig 2:PRISMA Flowchart for Literature Review}
\end{figure}

The initial search returned 1935 records, of which 367 duplicates were removed. The remaining 1568 records were imported into ASReview for screening. This tool iteratively prioritised studies based on relevance, facilitating the identification of 353 candidate articles for full-text assessment. During this stage, a refined set of inclusion and exclusion criteria was applied to evaluate each study’s suitability.

Eligible studies met all of the following conditions: (1) they investigated cycling behaviour, perception, or experience in urban settings; (2) they included at least one spatial scale (macro, micro, or hybrid); (3) they applied empirically valid and interpretable methods; (4) and they were peer-reviewed journal articles written in English. Studies were excluded if they focused exclusively on sport or leisure cycling (e.g., mountain biking), rural or natural contexts, newly developing cycling systems, or if they were non-empirical in nature (e.g., commentaries or reviews).

Following full-text review, 292 studies were excluded based on these criteria. An additional 10 records were identified through reference screening and citation tracking. A total of 68 studies were ultimately included in the review (Figure \ref{fig 2:PRISMA Flowchart for Literature Review}).

\subsection{Data extraction}

Key information was extracted from each of the 68 included studies. This included basic metadata such as title, author, publication year, and journal, as well as core analytical attributes relevant to our review. These attributes covered: (1) the spatial scale of analysis (macro, micro, or hybrid); (2) methodological approaches (objective, subjective, qualitative, or quantitative); (3) built environment determinants examined; and (4) any specification of demographic targeting (e.g., children, women, older adults).

\subsection{Taxonomy and thematic clusters}

\begin{figure}[ht]
\centering
\includegraphics[width=1 \linewidth, trim={0cm 0cm 10cm 0cm},clip]{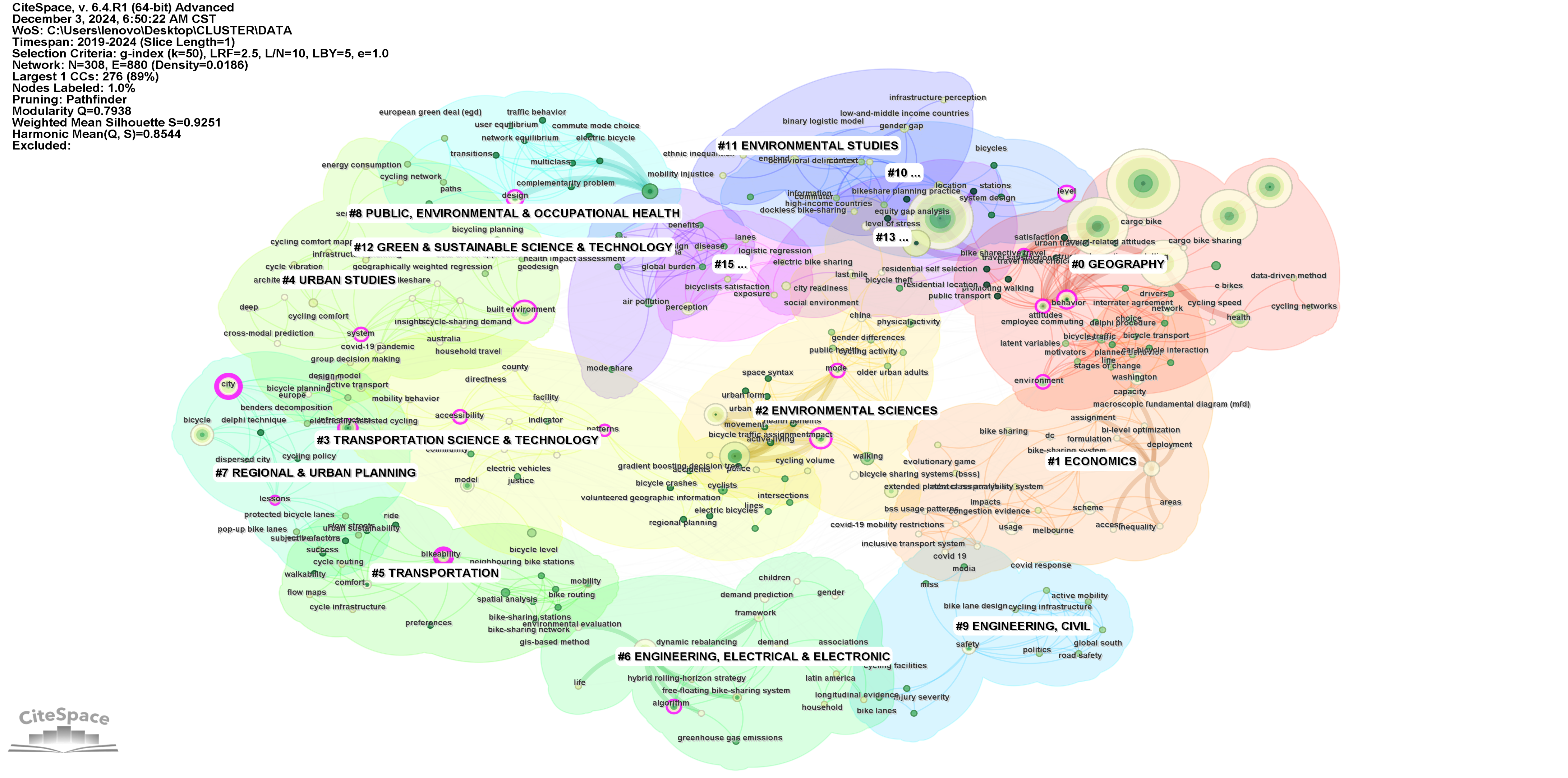}
\caption{Clusters of Disciplines (Software: CiteSpace)}
\label{fig 4:Clusters of Disciplines}
\end{figure}

After examining all relevant studies, we established a thematic taxonomy to guide subsequent analysis. Disciplinary clusters were identified using CiteSpace \citep{chen_searching_2004}, based on co-citation and keyword proximity (Figure~\ref{fig 4:Clusters of Disciplines}). The results reveal that most research emerges from transportation, urban planning, and environmental sciences, with behavioural studies and public health forming smaller but recognisable groups. These disciplinary orientations inform how built environment variables are framed: transportation studies emphasise infrastructural systems at the city scale, whereas urban planning and environmental psychology focus more on street-level design and perceptual qualities. This taxonomy forms the analytic foundation for subsequent classification and interpretation.

\section{Results}

\begin{figure}[ht] 
    \centering
    \includegraphics[width=1\linewidth]{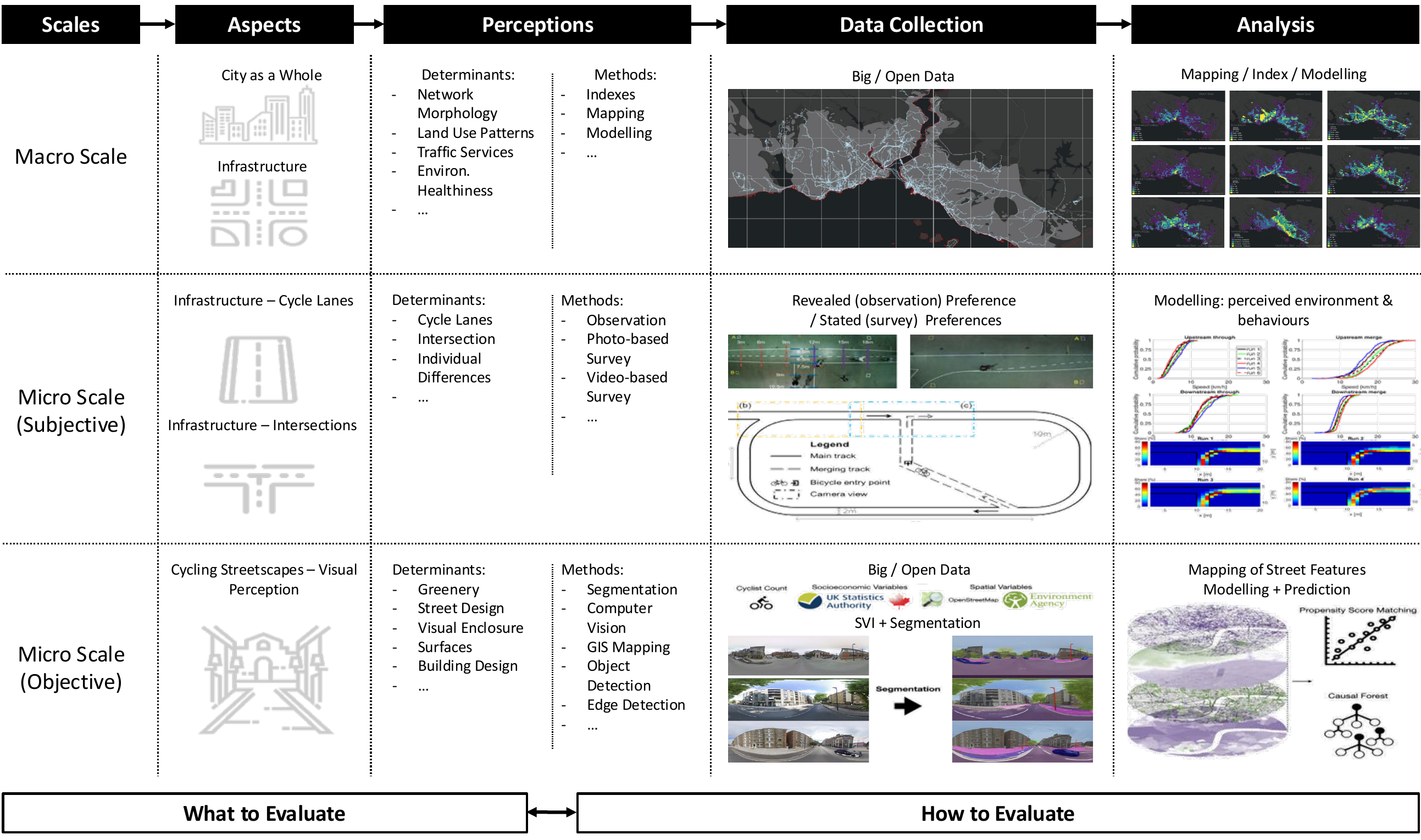} 
    \caption{Typical Low of Research. \newline Credit: The icons used in this diagram are obtained by the Noun Project, and the example studies of macro, micro, and hybrid scales are \citet{ciris_investigating_2024}, \citet{gavriilidou_empirical_2021}, and \citet{fujiwara_panorama-based_2024} accordingly.} 
    \label{fig 3:Typical Low of Research} 
\end{figure}

Reviewed studies demonstrate a clear methodological convergence despite spanning multiple disciplines and spatial scales. As visualised in Figure~\ref{fig 3:Typical Low of Research}, most analyses follow a four-step structure: (1) identifying aspects of the built environment to examine (e.g., network connectivity, infrastructure, and cycling streetscapes); (2) Identifying aspects of cycling experience and perceptions (e.g., diversity, safety, and attractiveness); (3) identifying specific data (e.g., big geo data and street view imagery); (4) analysing the collected data. Despite disciplinary variation, this structure recurs across urban scales, indicating a common analytical logic that frames much of the current research \citep{ito_understanding_2024}.

\begin{figure}[ht] 
    \centering
    \begin{subfigure}[t]{0.48\linewidth}
    \includegraphics[width=1\linewidth]{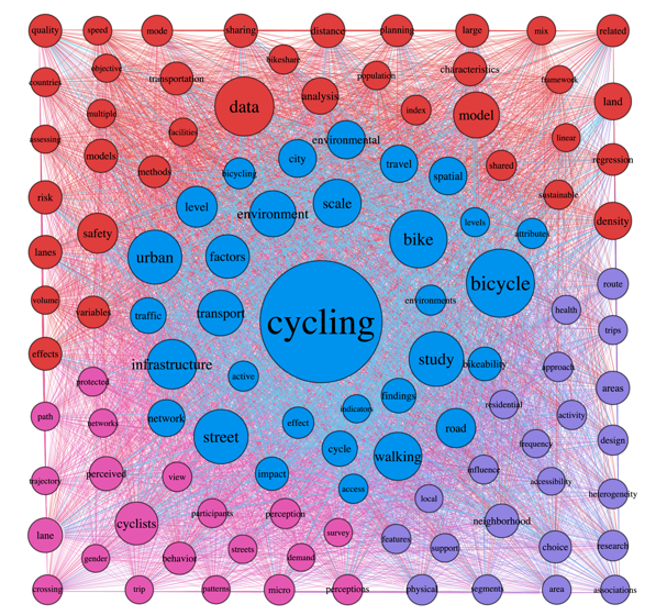} 
    \caption{Network Analysis of Keywords} 
    \label{fig 6:Network Analysis of Keywords} 
    \end{subfigure}
    \hfill
    \begin{subfigure}[t]{0.48\linewidth}
    \centering
    \includegraphics[width=1\linewidth]{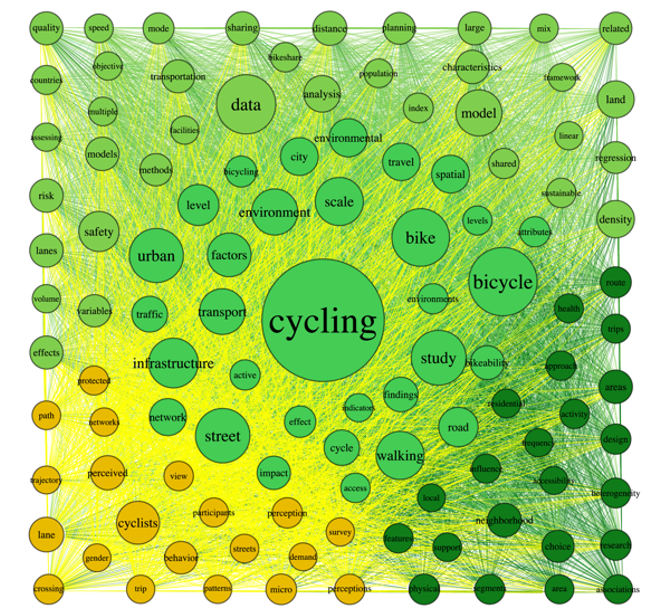} 
    \caption{Chronological Distributions of Keywords} 
    \label{fig 7:Chronological Distributions of Keywords} 
    \end{subfigure}
    \caption{Network Analysis of Keywords (Software: GEPHI)}
\end{figure}

Keyword clustering reveals a threefold thematic distribution structured by scale and disciplinary orientation. In Figure~\ref{fig 6:Network Analysis of Keywords}, macro-scale studies form the upper cluster, linking cycling infrastructure with urban mobility and transport policy. Neighbourhood-level analyses occupy the lower right, with emphasis on route choice and accessibility. The lower left cluster reflects a distinct strand of research focused on micro-scale perceptual attributes, such as visibility, legibility, and streetscape aesthetics. These clusters suggest a shifting focus: while transport-centred models remain dominant, there is increasing engagement with street-level experience and perceptual legibility.

Temporal dynamics reinforce this thematic transition. As shown in Figure~\ref{fig 7:Chronological Distributions of Keywords}, neighbourhood-scale studies peaked between 2014 and 2018, primarily addressing accessibility and travel behaviour through infrastructure-based metrics. By contrast, studies foregrounding subjective perception, street view imagery, and machine learning techniques have accelerated since 2022, reflecting both methodological innovation and growing interest in the experiential dimensions of cycling.

\begin{figure}[ht] 
    \centering
    \includegraphics[width=0.7\linewidth]{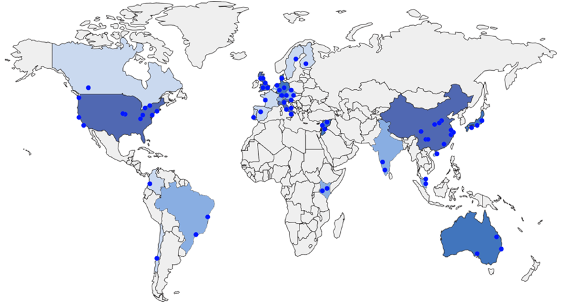} 
    \caption{Geographic Distributions of Study Areas (Basemap: carto)} 
    \label{fig 8:Network Analysis of Keywords} 
\end{figure}

Geographic distribution further contextualises this shift. As illustrated in Figure~\ref{fig 8:Network Analysis of Keywords}, a substantial proportion of studies are situated in European and East Asian cities, where historical morphology presents both constraints and opportunities for cycling. These environments offer a natural testbed for examining fine-grained interventions. In North American and Australian cities, by contrast, studies tend to focus on large-scale retrofitting and infrastructure expansion. Across all regions, there is a notable preference for urban cores and high-density residential areas.

\begin{figure}[ht] 
    \centering
    \includegraphics[width=1\linewidth]{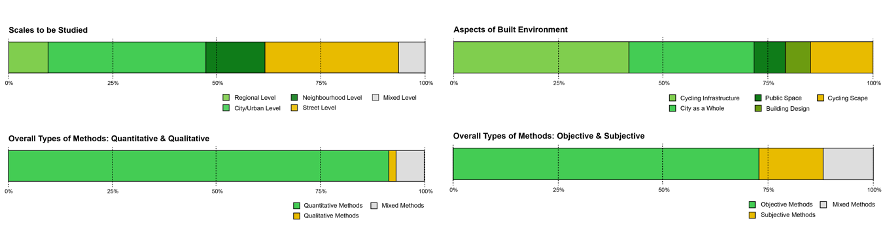} 
    \caption{Bar Plots, the figure portrays: (top left) scales to be studied. (top right) aspects of built environment. (bottom left) overall types of methods: quantitative and qualitative. (bottom right) overall types of methods: objective and subjective.} 
    \label{fig 9:Bar Plots} 
\end{figure}

Overall methodological tendencies remain skewed toward quantitative and objective approaches. Figure~\ref{fig 9:Bar Plots} shows that most studies operate at the city or street level (top left), with cycling infrastructure and general urban configuration as dominant environmental foci (top right). Quantitative methods are used more frequently than qualitative ones, and objective metrics are prioritised over subjective perception (bottom plots). Nonetheless, the increasing presence of user-reported data and perceptual evaluation suggests a gradual rebalancing between form-based and experience-based perspectives.

\clearpage
\newpage
\section{Review}
\subsection{The ‘ability to cycle’ at the macro scale}
\begin{table}[htbp]
    \centering
    \renewcommand{\arraystretch}{2}
    \setlength{\tabcolsep}{5pt} 
    \resizebox{\textwidth}{!}{ 
    \begin{tabular}{ p{4.5cm} p{1.5cm} p{3 cm} p{1cm} c c c c c c c c c c c c c }
        \toprule
        & \multicolumn{3}{c}{\textbf{Environment}} & \multicolumn{4}{c}{\textbf{Data}} & \multicolumn{3}{c}{\textbf{Methods}} & \multicolumn{6}{c}{\textbf{Determinants}} \\
        \cmidrule(lr){2-4} \cmidrule(lr){5-8} \cmidrule(lr){9-11} \cmidrule(lr){12-17}
        
        \rotatebox{90}{} & 
        \rotatebox{90}{\textbf{Study Scale}} & 
        \rotatebox{90}{\textbf{Urban Contexts}} & 
        \rotatebox{90}{\textbf{Agents}} &
        \rotatebox{90}{\textbf{Surveys}} & 
        \rotatebox{90}{\textbf{Observations}} & 
        \rotatebox{90}{\textbf{Open/Big Data}} & 
        \rotatebox{90}{\textbf{Sample Size}} &
        \rotatebox{90}{\textbf{Quantitative/Qualitative}} &
        \rotatebox{90}{\textbf{Objective/Subjective}} & 
        \rotatebox{90}{\textbf{Metrics/Methods}} &
        \rotatebox{90}{\textbf{Network Connectivity}} & 
        \rotatebox{90}{\textbf{Land Use Patterns}} &
        \rotatebox{90}{\textbf{Traffic Services}} & 
        \rotatebox{90}{\textbf{Environmental Healthiness}} &
        \rotatebox{90}{\textbf{Street Infrastructure}} & 
        \rotatebox{90}{\textbf{Cycling Streetscapes}} \\ 
        
        \midrule
        \makecell[l]{Ciris, Akay and Tumer, 2024} & City & Istanbul & bike & \ding{55} & \ding{55} & \ding{51} & \ding{55} & q & o & MGWR & \ding{51} & \ding{51} & \ding{51} & \ding{55} & \ding{55} & \ding{55} \\
        
        \makecell[l]{Huardinghaus et al., 2021} & City & Berlin & bike & \ding{51} & \ding{55} & \ding{51} & \ding{55} & q & o & Bikeability INDEX & \ding{51} & \ding{51} & \ding{51} & \ding{55} & \ding{55} & \ding{55} \\
        
        \makecell[l]{Codina et al., 2022} & City & Barcelona & bike & \ding{55} & \ding{55} & \ding{51} & \ding{55} & q & o & Bikeability INDEX & \ding{51} & \ding{51} & \ding{51} & \ding{55} & \ding{55} & \ding{55} \\
        
        \makecell[l]{Gao et al., 2021} & City & Shanghai & Shared bike & \ding{55} & \ding{55} & \ding{51} & \ding{55} & q & o & MGWR; GIS & \ding{51} & \ding{51} & \ding{51} & \ding{55} & \ding{55} & \ding{55} \\

        \makecell[l]{Beura et al., 2021} & City & Indian Cities & bike & \ding{55} & \ding{55} & \ding{51} & \ding{55} & q & o & GPS & \ding{55} & \ding{51} & \ding{55} & \ding{51} & \ding{55} & \ding{55} \\
        
        \makecell[l]{Gong, Rui and Li, 2024} & City & New York & Shared bike & \ding{55} & \ding{51} & \ding{51} & \ding{55} & q & o & MGWR & \ding{51} & \ding{51} & \ding{51} & \ding{51} & \ding{55} & \ding{55} \\
        
        \makecell[l]{Boakye et al., 2023} & Commu -nity & US communities & bike & \ding{51} & \ding{55} & \ding{55} & 39.908 & l & s & NEWS & \ding{51} & \ding{51} & \ding{51} & \ding{51} & \ding{55} & \ding{55} \\

        \makecell[l]{Ferrari et al., 2020} & Neighbour -hood & Latin American & bike & \ding{51} & \ding{51} & \ding{51} & 9,218 & l & s & NEWS & \ding{51} & \ding{51} & \ding{51} & \ding{51} & \ding{55} & \ding{55} \\
        
        \bottomrule
        
        \multicolumn{17}{c}{\textbf{Symbols:}\hspace{10 cm} \ding{51}: Feature, \ding{55}: Not Feature, q: quantitative, l: qualitative, o: objective, s: subjective.}
    \end{tabular}}
    \caption{Overview of studies investigating cycling experience in macro-scale historical urban contexts. \newline Studies are organised into four main categorises: environment (study scale; urban contexts; and agent of study), Data Collection (surveys; observations; open/big data; sample size), methodology (quantitative; qualitative; objective; subjective; metrics/methods), and determinants (network morphology; land use patterns; traffic services; environmental healthiness; street infrastructure; cycling streetscapes).}
    \label{tab 1:Macro Scale}
\end{table}

Research situated at the macro scale consistently frames cycling not as an isolated act of movement, but as a product of spatial systems that condition its feasibility. These systems, spanning regional to neighbourhood scales, define what is structurally possible for cyclists through the configuration of routes, densities, and proximities. Within this framing, the ‘5Ds’ model (i.e., density, diversity, destination accessibility, distance to transit, and design) has gained prominence as a conceptual scaffold for identifying structural determinants of cycling experience \citep{wang_designing_2023}. The studies summarised in Table~\ref{tab 1:Macro Scale} exemplify this spatial emphasis.

Among the most consequential determinants is network morphology, which functions not merely as background infrastructure but as a spatial logic that structures movement. A consistent body of work identifies street connectivity (measured through intersection density, block size, and directional coherence) as a predictor of route continuity and cyclist uptake \citep{chen_quantify_2025, schon_scoping_2024, wang_relationships_2023, brauer_characterizing_2021, ding_towards_2021, rupi_data-driven_2019, liu_development_2020}. Fine-grained grids and reduced block lengths are shown to instantiate legible and uninterrupted trajectories, thereby reducing friction and enhancing perceived safety \citep{sun_examining_2017, cunha_assessing_2023, gao_unraveling_2023, beura_bicycle_2021}. These morphological affordances are frequently reinforced by compact land-use patterns and multimodal integration, particularly where access to public transport and urban density coalesce \citep{ciris_investigating_2024, hardinghaus_more_2021, nasri_analysis_2020, hamilton_bicycle_2018, la_paix_role_2021}.

Environmental exposure introduces an additional scalar layer of influence. Air quality, noise, precipitation, and wind conditions exert non-trivial effects on cycling frequency and user preference, often by amplifying latent infrastructural shortcomings \citep{codina_built_2022, gong_deciphering_2024, zhou_assessing_2023, ciris_investigating_2024}. While such factors are typically treated as exogenous, several studies incorporate them into integrative models, where environmental constraints intersect with urban form. At the neighbourhood scale, macro-scale studies also recognise the role of perception, especially in relation to built density, land use mix, and access networks (commonly summarised as the ‘DMA’ triad) \citep{ferrari_association_2020, karmeniemi_residential_2019}. These subjective qualities often align with objective indicators, reinforcing the multi-layered nature of environmental legibility.

The methodological landscape of macro-scale research is shaped by spatial analytics and computational generalisation. GIS-based models and Space Syntax remain central to tracing how structural configurations shape cycling flows \citep{codina_built_2022, deng_multi-scale_2024}. These tools are routinely paired with aggregated spatial datasets (including census records, GPS traces, and POI distributions) to generate composite indices such as the Bikeability Index, the Neighbourhood Environment Walkability Survey (NEWS), and Multi-Scale Geographically Weighted Regression (MGWR) \citep{hamilton_bicycle_2018, song_unraveling_2024, rupi_data-driven_2019, gao_spatial_2021, yang_key_2023, yuan_cycling_2024, ni_evaluation_2024}. While such frameworks enable generalisability across urban contexts, they also introduce dependencies on data quality and spatial resolution, limiting portability across regions \citep{hardinghaus_more_2021, li_assessing_2025}. Although some studies integrate self-assessed perceptions and field-based observations, these remain peripheral to the dominant modelling paradigm \citep{kerr_perceived_2016, piras_perceived_2023}.

\subsection{The ‘propensity to cycle’ at the micro scale}

\begin{table}[hb]
    \centering
    \small
    \renewcommand{\arraystretch}{2}
    \setlength{\tabcolsep}{5pt} 
    \resizebox{\textwidth}{!}{ 
    \begin{tabular}{ p{4.5cm} p{1.5cm} p{3 cm} p{1cm} c c c c c c c c c c c c c }
        \toprule
        & \multicolumn{3}{c}{\textbf{Environment}} & \multicolumn{4}{c}{\textbf{Data}} & \multicolumn{3}{c}{\textbf{Methods}} & \multicolumn{6}{c}{\textbf{Determinants}} \\
        \cmidrule(lr){2-4} \cmidrule(lr){5-8} \cmidrule(lr){9-11} \cmidrule(lr){12-17}
        
        \rotatebox{90}{} & 
        \rotatebox{90}{\textbf{Study Scale}} & 
        \rotatebox{90}{\textbf{Urban Contexts}} & 
        \rotatebox{90}{\textbf{Agents}} &
        \rotatebox{90}{\textbf{Surveys}} & 
        \rotatebox{90}{\textbf{Observations}} & 
        \rotatebox{90}{\textbf{Open/Big Data}} & 
        \rotatebox{90}{\textbf{Sample Size}} &
        \rotatebox{90}{\textbf{Quantitative/Qualitative}} &
        \rotatebox{90}{\textbf{Objective/Subjective}} & 
        \rotatebox{90}{\textbf{Metrics/Methods}} &
        \rotatebox{90}{\textbf{Network Connectivity}} & 
        \rotatebox{90}{\textbf{Land Use Patterns}} &
        \rotatebox{90}{\textbf{Traffic Services}} & 
        \rotatebox{90}{\textbf{Environmental Healthiness}} &
        \rotatebox{90}{\textbf{Street Infrastructure}} & 
        \rotatebox{90}{\textbf{Cycling Streetscapes}} \\ 
        
        \midrule
        \makecell[l]{Bialkova et al., 2022} & Street & Dutch City & bike & \ding{51} & \ding{51} & \ding{55} & 77 & q & s & VR; attractiveness & \ding{55} & \ding{55} & \ding{55} & \ding{55} & \ding{51} & \ding{55} \\
        
        \makecell[l]{Bogacz et al., 2021} & Street & Singapore & bike & \ding{51} & \ding{51} & \ding{55} & 48 & q & s & VR; risk perception & \ding{55} & \ding{55} & \ding{55} & \ding{55} & \ding{51} & \ding{55} \\
        
        \makecell[l]{Nazemi et al., 2021} & Street & Singapore & bike & \ding{51} & \ding{51} & \ding{55} & 150 & q & s & VR; PLOS; WTB & \ding{55} & \ding{55} & \ding{55} & \ding{55} & \ding{51} & \ding{55} \\
        
        \makecell[l]{Guo et al., 2021} & Street & Charlottesville & bike & \ding{51} & \ding{51} & \ding{55} & 50 & q & s & VR; eye tracking; HR & \ding{55} & \ding{55} & \ding{55} & \ding{55} & \ding{51} & \ding{55} \\

        \makecell[l]{Mertens et al., 2016} & Street & Collaged Photos & bike & \ding{51} & \ding{51} & \ding{55} & 1,950 & l & s & GPS & \ding{55} & \ding{55} & \ding{55} & \ding{55} & \ding{51} & \ding{55} \\
        
        \makecell[l]{Gavriilidou et al., 2021} & Street & Experient Setup & e-bike; bike & \ding{51} & \ding{51} & \ding{51} & 88 & q & s & real-world experiment & \ding{55} & \ding{55} & \ding{55} & \ding{55} & \ding{51} & \ding{55} \\
                
        \bottomrule
        
        \multicolumn{17}{c}{\textbf{Symbols:}\hspace{10 cm} \ding{51}: Feature, \ding{55}: Not Feature, q: quantitative, l: qualitative, o: objective, s: subjective.}
    \end{tabular}
    }
    \caption{Overview of studies investigating cycling experience in micro-scale historical urban contexts.}
    \label{tab 2:Micro Scale}
\end{table}

In contrast to macro-scale studies that foreground structural accessibility, micro-scale research investigates cycling as a situated and interpretive experience that unfolds along street segments and intersections. This body of work conceptualises ‘propensity to cycle’ not as infrastructural provision alone, but as a function of how individuals perceive, evaluate, and physiologically engage with their immediate environment. Central constructs such as risk perception, Perceived Level of Stress (PLOS), and Willingness to Bicycle (WTB) are frequently operationalised to capture this orientation \citep{weikl_data-driven_2023, beirens_which_2024}. Streets, as the primary sites of embodied cycling experience, remain the empirical focus for studies situated at this scale \citep{ye_visual_2019, wang_relationships_2023, ahmed_bicycle_2024}. Table~\ref{tab 2:Micro Scale} provides an overview of representative studies.

Demographic variation mediates how infrastructural configurations are interpreted and acted upon. Certain populations (e.g., women, older adults, and children) exhibit consistently lower thresholds for perceived safety and tend to prefer physically segregated facilities \citep{aldred_diversifying_2017, basil_exploring_2023, battiston_revealing_2023}. Meanwhile, lower-income individuals often engage cycling as a functional necessity, placing greater emphasis on modal flexibility and affordability. To account for such heterogeneity, several studies adopt typological classifications that map users onto behavioural archetypes, including ‘strong and fearless’, ‘enthused and confident’, ‘interested but concerned’, and ‘no way, no how’ \citep{rerat_build_2024, dill_four_2013}. These categories are analytically useful in highlighting that infrastructural effectiveness is contingent upon user disposition, not merely physical form \citep{biassoni_choosing_2023}.

The design of street-level infrastructure exerts a direct influence on perceptual and affective responses during cycling. Physically protected lanes are repeatedly shown to reduce anticipatory stress and enhance perceived control, whereas complex or poorly signposted intersections increase cognitive and collision-related risks \citep{bialkova_how_2022, ahmed_assessing_2024}. Spatial variables such as lane width, buffer zones, and separation from motor traffic are found to mediate user satisfaction and behavioural predictability \citep{von_stulpnagel_how_2022, bialkova_how_2022}. Beyond functional legibility, visual cues embedded in the urban fabric (e.g., greenery, water bodies, or architectural articulation) contribute to the aesthetic and emotional valence of the cycling journey \citep{ahmed_bicycle_2024, li_assessing_2025}.

Empirical methods deployed in micro-scale studies prioritise subjective input, often categorised into stated preference (SP) and revealed preference (RP) approaches. SP methods rely on participants’ responses to hypothetical scenarios, typically presented through textual descriptions, static images, or pre-recorded video segments \citep{nazemi_studying_2021, mertens_which_2016}. RP methods, in contrast, observe real-time decisions in situ, using GPS traces or behavioural monitoring to derive patterns of route selection and engagement \citep{guo_psycho-physiological_2023, rerat_build_2024}. While SP approaches may be limited by imaginative distortion and insufficient realism, RP approaches often lack environmental control and are constrained by contextual unpredictability \citep{guo_psycho-physiological_2023, ghekiere_assessing_2015}.

Recent work has begun to incorporate immersive technologies to navigate between these methodological limitations. Virtual Reality (VR) simulations offer controlled yet high-fidelity representations of urban environments, enabling participants to engage with spatial variables under near-naturalistic conditions \citep{nazemi_studying_2021, guo_psycho-physiological_2023, beura_development_2022}. These setups mitigate physical risk while maintaining ecological relevance. When paired with biometric measurements (e.g., eye-tracking or heart rate monitoring) VR allows researchers to extract quantifiable indicators of stress, attention, and emotional state in relation to specific design features \citep{bialkova_how_2022, bogacz_cycling_2021}. However, the procedural demands of scenario construction remain high, and existing applications have thus far been largely confined to a limited typology of cycling environments, particularly intersections and protected lanes.

\subsection{Hybrid approach across methods and scales}

Recent studies have advanced hybrid approaches that integrate macro-scale urban morphology with micro-scale streetscape perceptions, providing a more nuanced understanding of how built environments shape cycling experience. These studies move beyond conventional scalar separations by situating large-scale indicators within fine-grained spatial contexts, thereby enabling cross-scale evaluation of street-level environments. At the core of this hybrid turn is the use of enriched spatial data sources, including street-view imagery (SVI), mobile trajectories, bike-sharing records, and deep-learning-assisted feature segmentation \citep{staves_modelling_2025, zare_simple_2024, garrido-valenzuela_where_2023, guo_moderation_2024, lu_using_2019, song_unraveling_2024, wang_relationship_2020}. Supplementary spatial metrics such as the Normalized Difference Vegetation Index (NDVI) and Points of Interest (POI) density are often incorporated to assess the functional, aesthetic, and perceptual characteristics of street segments \citep{ye_visual_2019, hankey_predicting_2021}. Table~\ref{tab 3:Hybrid Scale} presents an overview of studies employing hybrid-scale analytical strategies.

Streetscape-related attributes remain central to hybrid-scale assessments. Greenery is the most frequently analysed feature, consistently associated with increased comfort, reduced stress, and heightened willingness to cycle \citep{wang_relationship_2020, lu_associations_2019}. Complementary visual components, such as street-facing façades and sky openness, are found to influence perceptions of vitality and safety, particularly when interacting with broader indicators of land use diversity and public access \citep{zare_simple_2024, guo_multiple_2024, wang_designing_2023, ye_visual_2019}. Rather than treating these variables as additive, hybrid studies often model them as interdependent, foregrounding the ways in which fine-scale features modulate the influence of macro-level urban conditions.

Methodological advancements underpinning this hybridisation are primarily data-driven. The increasing availability of geolocated image datasets, coupled with computational segmentation algorithms, allows for the scalable extraction of visual indicators across large networks. Deep learning techniques now enable the automated classification of streetscape features, significantly reducing the resource demands of field-based audits \citep{staves_modelling_2025, ito_examining_2024, meng_street_2021, qin_investigating_2024, zhang_encouraging_2024}. In addition, multi-source data integration—linking SVI-derived metrics with behavioural traces or population-level travel logs—permits more robust inference across temporal and spatial scales. Nonetheless, these approaches remain constrained by data unevenness across geographies and the interpretive limitations of proxy variables when assessing subjective experience.

\begin{table}[ht]
    \centering
    \small
    \renewcommand{\arraystretch}{2}
    \setlength{\tabcolsep}{5pt} 
    \resizebox{\textwidth}{!}{ 
    \begin{tabular}{ p{4.5cm} p{1.5cm} p{3 cm} p{1cm} c c c c c c c c c c c c c }
        \toprule
        & \multicolumn{3}{c}{\textbf{Environment}} & \multicolumn{4}{c}{\textbf{Data}} & \multicolumn{3}{c}{\textbf{Methods}} & \multicolumn{6}{c}{\textbf{Determinants}} \\
        \cmidrule(lr){2-4} \cmidrule(lr){5-8} \cmidrule(lr){9-11} \cmidrule(lr){12-17}
        
        \rotatebox{90}{} & 
        \rotatebox{90}{\textbf{Study Scale}} & 
        \rotatebox{90}{\textbf{Urban Contexts}} & 
        \rotatebox{90}{\textbf{Agents}} &
        \rotatebox{90}{\textbf{Surveys}} & 
        \rotatebox{90}{\textbf{Observations}} & 
        \rotatebox{90}{\textbf{Open/Big Data}} & 
        \rotatebox{90}{\textbf{Sample Size}} &
        \rotatebox{90}{\textbf{Quantitative/Qualitative}} &
        \rotatebox{90}{\textbf{Objective/Subjective}} & 
        \rotatebox{90}{\textbf{Metrics/Methods}} &
        \rotatebox{90}{\textbf{Network Connectivity}} & 
        \rotatebox{90}{\textbf{Land Use Patterns}} &
        \rotatebox{90}{\textbf{Traffic Services}} & 
        \rotatebox{90}{\textbf{Environmental Healthiness}} &
        \rotatebox{90}{\textbf{Street Infrastructure}} & 
        \rotatebox{90}{\textbf{Cycling Streetscapes}} \\ 
        
        \midrule
        \makecell[l]{Hankey et al., 2021} & Street & US cities & bike & \ding{55} & \ding{55} & \ding{51} & \ding{55} & q & o & SVI; POI; Census data & \ding{55} & \ding{55} & \ding{55} & \ding{55} & \ding{51} & \ding{51} \\
        
        \makecell[l]{Ito, Bansal and Bijecki, 2024} & Street & London & bike & \ding{55} & \ding{55} & \ding{51} & \ding{55} & q & o & SVI; segmentation & \ding{55} & \ding{55} & \ding{55} & \ding{55} & \ding{51} & \ding{55} \\
        
        \makecell[l]{Lu et al., 2019} & Street & Hong Kong & bike & \ding{51} & \ding{55} & \ding{51} & 5,701 & q & o & NDVI; SVI; Census data & \ding{55} & \ding{55} & \ding{55} & \ding{55} & \ding{55} & \ding{51} \\
        
        \makecell[l]{Guo et al., 2024} & Street & Shanghai & Shared bike & \ding{55} & \ding{55} & \ding{51} & \ding{55} & q & s & SVI; bike-sharing data & \ding{55} & \ding{55} & \ding{55} & \ding{55} & \ding{51} & \ding{51} \\

        \makecell[l]{Song et al., 2024} & Street & Ithaca & Shared bike & \ding{55} & \ding{55} & \ding{51} & \ding{55} & q & o & SVI; bike-sharing data & \ding{55} & \ding{55} & \ding{55} & \ding{55} & \ding{51} & \ding{51} \\
        
        \makecell[l]{Wang et al., 2020} & Street & Shenzhen & SHared bike & \ding{55} & \ding{55} & \ding{51} & \ding{55} & q & o & SVI; bike-sharing data & \ding{55} & \ding{55} & \ding{55} & \ding{55} & \ding{51} & \ding{51} \\
                
        \bottomrule
        
        \multicolumn{17}{c}{\textbf{Symbols:}\hspace{10 cm} \ding{51}: Feature, \ding{55}: Not Feature, q: quantitative, l: qualitative, o: objective, s: subjective.}
    \end{tabular}
    }
    \caption{Overview of studies investigating cycling experience using hybrid-scale analysis.}
    \label{tab 3:Hybrid Scale}
\end{table}

\subsection{Cycling experience in historical contexts: the case of Cambridge}

\begin{figure}[htbp] 
    \centering
    \includegraphics[width= 1 \linewidth]{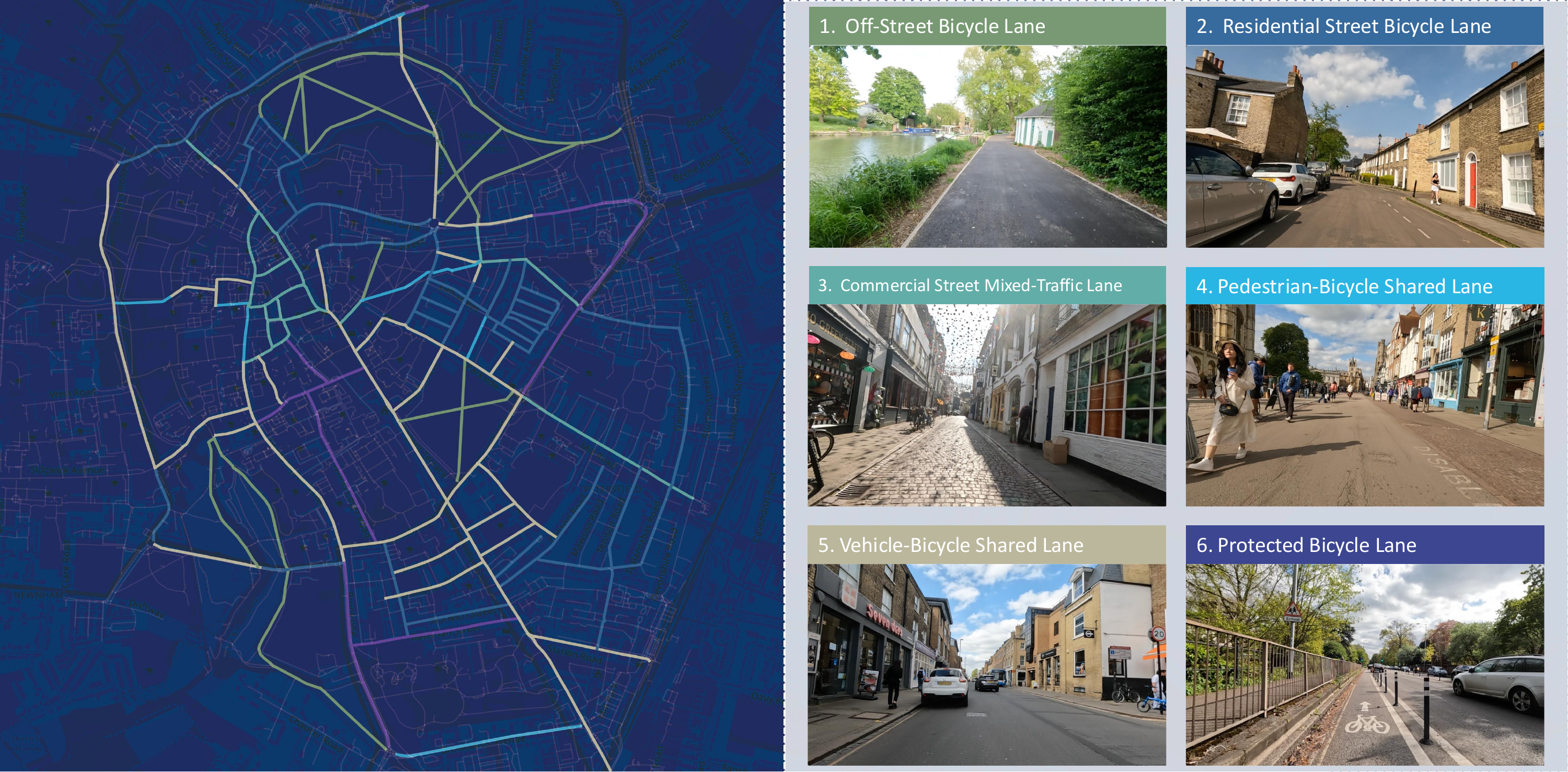} 
    \caption{Study Area and Street Typologies}
    \label{fig 2.1-2} 
\end{figure}

Historical urban contexts pose distinctive challenges and opportunities for cycling, often shaped by inherited spatial morphologies, fine-grain networks, and embedded heritage value. These settings typically exhibit pre-modern street grids, compact urban fabrics, and limited right-of-way widths, all of which shape how contemporary mobility infrastructures can be retrofitted. Cambridge exemplifies such conditions: a medieval city core where cultural continuity and spatial constraints co-exist. As a city with the highest cycling rates in the United Kingdom, Cambridge has been consistently referenced in cycling studies for its integration of land use, climate moderation, and civic cycling culture, often described as a “cycling enclave” within a motorised national context \citep{aldred_outside_2010, carse_factors_2013}. However, its preserved historical morphology also reveals inherent frictions between spatial inheritance and infrastructural adaptation.

In this review, the historical city centre of Cambridge is taken as a representative case to explore cycling experience under morphologically constrained conditions. Field-based investigations were conducted by the authors, including real-time cycling, visual documentation via GoPro-mounted recordings, and systematic segment classification. To account for spatial variation, 116 recorded street segments were grouped into six typologies  (Figure \ref{fig 2.1-2}) adapted from established schemas \citep{houde_ride_2018, nazemi_studying_2021, noland_understanding_2023, teschke_proximity_2017, zhang_measuring_2022}: (1) off-street bicycle lanes, (2) residential street bicycle lanes, (3) commercial street mixed-traffic lanes, (4) pedestrian–bicycle shared lanes, (5) vehicle–bicycle shared lanes, and (6) protected bicycle lanes. Each typology reflects a distinctive spatial configuration and traffic interaction mode, offering a structured lens for analysing cycling experience in a historically layered context.

\begin{figure}[ht] 
    \centering
    \includegraphics[width=0.8 \linewidth]{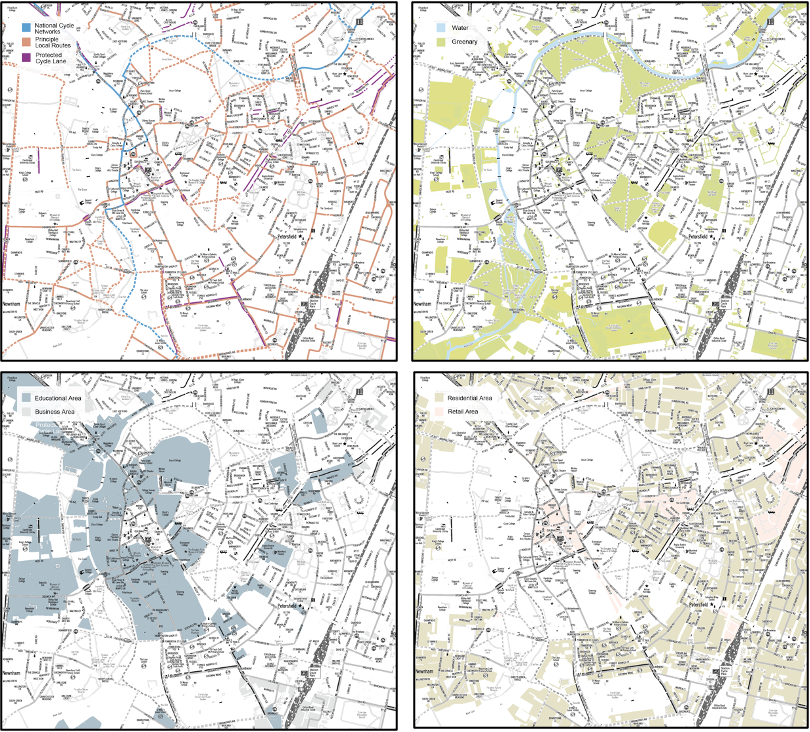} 
    \caption{ (top left) Cycling Networks in Cambridge. (top right) Land Use Patterns - Water and Greenery. (bottom left) Land Use Patterns – Educational and Business Areas. (bottom right) Residential and Retail Areas.} 
    \label{fig 10: (top left) Cycling Networks in Cambridge. (top right) Land Use Patterns - Water and Greenery. (bottom left) Land Use Patterns – Educational and Business Areas. (bottom right) Residential and Retail Areas.} 
\end{figure}

At the macro scale, the city’s compact structure facilitates a high degree of functional connectivity. Land use is tightly integrated, with educational institutions, commercial areas, and residential zones located in close proximity. Such spatial integration reduces average trip distances and reinforces active mode choices (Figure~\ref{fig 10: (top left) Cycling Networks in Cambridge. (top right) Land Use Patterns - Water and Greenery. (bottom left) Land Use Patterns – Educational and Business Areas. (bottom right) Residential and Retail Areas.}). However, network redundancy is often limited by the historic layout, producing conditions in which cyclists must share space with vehicular traffic or traverse constrained intersections. These spatial limitations highlight the constraints of conventional infrastructural separation and underscore the importance of spatial legibility and perceptual continuity in shaping mobility experience.

\begin{figure}[ht] 
    \centering
    \includegraphics[width=0.8 \linewidth]{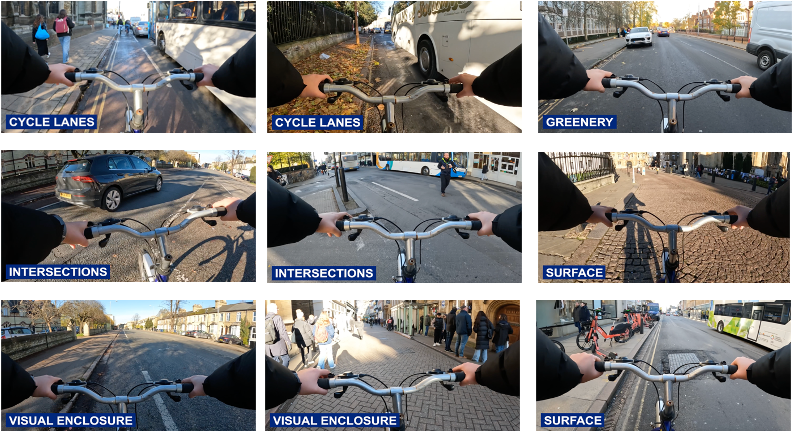} 
    \caption{ Street Views from Cyclists’ Perspective 
(Go-Pro Video: https://youtu.be/ZLDJns33LTQ?si=qnq8GPtzvoFiBCTV)} 
    \label{fig 11: Street Views from Cyclists’ Perspective} 
\end{figure}

At the micro scale, Cambridge’s streetscapes exhibit a diverse range of visual, material, and spatial features. First-person video documentation (Figure \ref{fig 11: Street Views from Cyclists’ Perspective}) revealed how continuous façades, canopy cover, surface materials, and enclosure influence perceptual outcomes. Positive impressions were associated with tree-lined corridors and spatial clarity, while discomfort was linked to uneven paving, narrow widths, and close contact with traffic. For novice cyclists, safety perceptions were especially sensitive to shared lanes and intersections. These findings suggest that the quality of cycling experience is not only infrastructural but also perceptual and cognitive, particularly in morphologically dense environments \citep{wang_relationship_2020, lu_research_2023}.

\FloatBarrier

\section{Conclusion and Discussion}
Across disciplinary and methodological domains, one finding remains consistent: environments shape cycling not only through spatial form but through how they are perceived, interpreted, and enacted. Infrastructure defines potential, but it is the experiential quality of the journey that converts capacity into behaviour.

This review identifies a two-scalar structure underlying existing research. At the macro level, connectivity, compactness, and mixed-use intensity enhance the ability to cycle by reinforcing legibility, coherence, and access. Environmental conditions (i.e., air quality, traffic flow, and climate) further modulate functional viability. At the micro level, the emphasis shifts toward propensity: how surface quality, delineation, greenery, and enclosure affect perceptions of comfort, safety, and appeal. These effects are not embedded in form alone but unfold through embodied experience.

Methodological tendencies align with this scalar distinction. Macro-scale studies privilege objective indicators derived from GIS, network metrics, and spatial analysis. Micro-scale research increasingly foregrounds subjective dimensions, capturing perception through surveys, experiments, and immersive simulation. Between these poles, hybrid approaches have begun to consolidate. By combining objective metrics with subjective evaluations, they offer a more integrated account of street-level environments. Rather than isolating infrastructure and perception, these methods acknowledge their mutual constitution, particularly salient in historical contexts where structural change is limited but perceptual cues remain actionable.

\subsection{Limitations}

\begin{figure}[ht] 
    \centering
    \begin{subfigure}{0.48\linewidth}
    \includegraphics[width=1\linewidth]{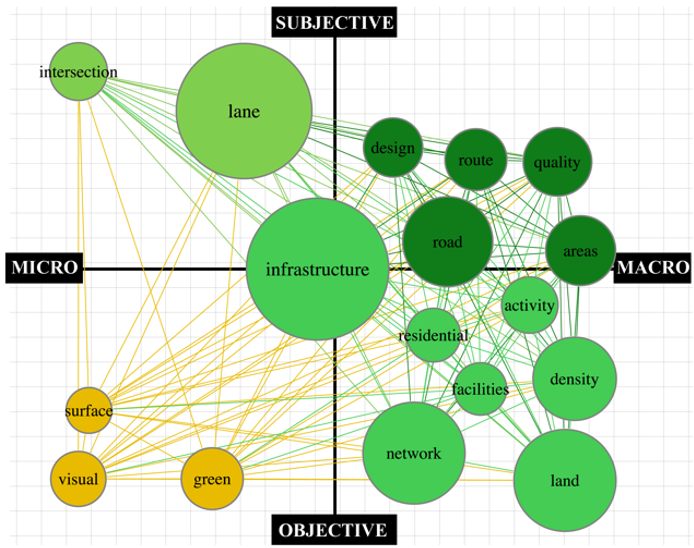} 
    \caption{Mapping of Determinants Discussed in Review} 
    \label{fig 13: Mapping of Determinants Discussed in Review} 
    \end{subfigure}
    \hfill
    \begin{subfigure}{0.48\linewidth}
    \centering
    \includegraphics[width=1\linewidth]{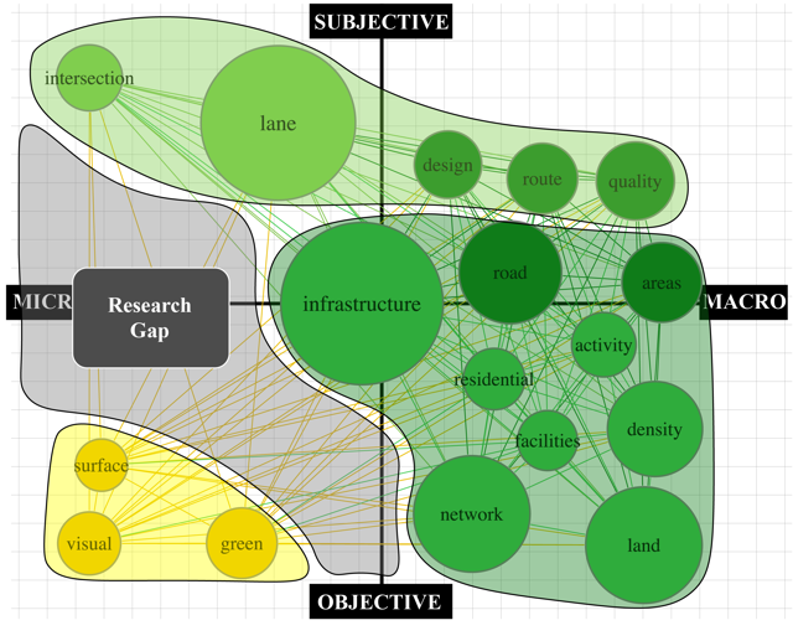} 
    \caption{Research Deficiencies and Gap} 
    \label{fig 14: Research Deficiencies and Gap} 
    \end{subfigure}
\end{figure}

Yet despite this emerging alignment between scale and method, several critical limitations persist. When mapped across a grid spanning macro to micro scales and objective to subjective approaches, the literature reveals significant asymmetries in methodological focus and empirical scope (Figure~\ref{fig 13: Mapping of Determinants Discussed in Review}).

Objective approaches dominate current research, particularly at the macro scale (see dark green zone in Figure~\ref{fig 14: Research Deficiencies and Gap}). These studies often rely on spatial analysis, network indices, and land-use typologies to infer cycling potential. While effective in modelling large-scale systems, they offer limited insight into the experiential quality of cycling. Key aspects such as perceived risk, navigational stress, or momentary discomfort remain unaccounted for \citep{ni_evaluation_2024}. Moreover, disparities in data availability and quality across regions hinder reproducibility and limit the external validity of model outputs \citep{guo_psycho-physiological_2023, hardinghaus_more_2021}.

Subjective approaches remain restricted in both methodological diversity and analytical resolution (see light green zone in Figure~\ref{fig 14: Research Deficiencies and Gap}). Most studies rely on stated or revealed preference surveys, which are prone to recall bias, hypothetical distortion, and spatial generalisation. Virtual reality experiments have been introduced to address these shortcomings, yet they often fall short of ecological validity. The sensory and embodied complexity of cycling through real urban environments --- affected by motion, attention, and affect --- is rarely captured. Consequently, subjective studies tend to focus on isolated micro-scale elements such as intersections or lane typologies, leaving broader compositional aspects of streetscapes insufficiently addressed.

A further limitation lies in the partial integration of street-level attributes into empirical models (see yellow zone in Figure~\ref{fig 14: Research Deficiencies and Gap}). Elements such as surface continuity, edge legibility, greenery, and visual enclosure substantially shape how cycling environments are experienced. However, these features are often assessed through computer vision and Street View Imagery (SVI), which privilege measurable appearance over embodied perception. Most implementations rely on static, panoramic imagery rather than dynamic, eye-level perspectives, which limits their capacity to reflect the real-time cues that influence comfort, safety, and wayfinding \citep{biljecki_street_2021}.

\subsection{Future research}

While research on the built environment and cycling experience has expanded across spatial and disciplinary boundaries, its methodological centre remains in flux. The rise of hybrid approaches marks a promising shift, but opportunities remain to consolidate and extend this direction.

At the street level, more work is needed to formalise how subjective evaluations relate to spatial features in ways that are generalisable and predictive \citep{ahmed_bicycle_2024, biassoni_choosing_2023, bialkova_how_2022, nielsen_bikeability_2018}. Many studies continue to treat perceptual feedback as contextual or anecdotal. Yet as computer vision and spatial analysis advance, there is scope to incorporate subjective scores into scalable models, enabling perceptual quality to be mapped, compared, and designed for \citep{yin_measuring_2016, qin_investigating_2024, li_quantifying_2018}.

A second opportunity lies in bridging descriptive analysis and design implementation \citep{pais_multicriteria_2022, cai_sidewalk-based_2024, mahfouz_road_2023}. While many studies identify perceptual determinants, few develop pathways for intervention. Linking subjective responses (e.g., discomfort or perceived safety) to modifiable features like greenery, edge continuity, or surface condition may guide interventions that remain sensitive to context. This is particularly relevant in constrained or heritage settings, where perceptual cues are often the most actionable \citep{carse_factors_2013, aldred_outside_2010}.

Finally, future research may benefit from expanding the methodological range of subjective assessment \citep{bretter_emotions_2025, bogacz_modelling_2021}. Physiological methods such as mobile EEG, GSR, or eye-tracking could offer continuous and implicit measures of affective and cognitive states \citep{abbas_neuroarchitecture_2024}. Immersive simulations using virtual or mixed reality may further support experimental control while approximating embodied motion. Together, these tools may allow subjective experience to be measured not only in retrospect, but in motion \citep{zeuwts_using_2023, zhao_responsive_2020, beirens_which_2024, ghanbari_use_2024}.










\bibliographystyle{cas-model2-names}

\bibliography{main}



\end{document}